\documentclass[twoside,leqno,twocolumn]{article}

\usepackage[letterpaper]{geometry}

\usepackage{siamproceedings}

\usepackage[T1]{fontenc}
\usepackage{amsfonts}
\usepackage{graphicx}
\usepackage{epstopdf}
\usepackage{enumitem}
\usepackage{algorithmic}
\ifpdf
  \DeclareGraphicsExtensions{.eps,.pdf,.png,.jpg}
\else
  \DeclareGraphicsExtensions{.eps}
\fi

\newsiamremark{remark}{Remark}
\newsiamremark{hypothesis}{Hypothesis}
\crefname{hypothesis}{Hypothesis}{Hypotheses}
\newsiamthm{claim}{Claim}

\usepackage{amsopn}

\newcommand\govt{}
\renewcommand\govt{\thanks{The U.S. Government retains a nonexclusive, royalty-free license to publish or reproduce the published form of this contribution, or allow others to do so, for U.S. Government purposes.}}

\begin{document}

\title{Application Failures and Machine Computational Efficiency\govt}

\author{Carlo Graziani \thanks{Argonne National Laboratory, Lemont, IL, USA (\email{cgraziani@anl.gov}, \email{blusch@anl.gov}).}
\and Bethany Lusch\footnotemark[1]
\and O. E. Bronson Messer \thanks{Oak Ridge National Laboratory, Oak Ridge, TN, USA (\email{bronson@ornl.gov})}}

\date{}

\maketitle

\fancyfoot[R]{\scriptsize{Copyright \textcopyright\ 2026\\
Copyright retained by principal author's organization}}

\begin{abstract}
We present a framework for evaluating uptime efficiency of Exascale-class scientific computers when application failure rates are appreciable. This is the situation that confronts current leadership-class scientific computing platforms and large AI training installations. What distinguishes scientific computing platforms is the heterogeneity of their applications. We argue that this diversity requires that failure rates and mean intervals between failures should be specified in terms of \emph{usage} (e.g. node-hours) rather than time, as is currently customary. We consider the usage loss terms due to failures, to checkpointing, and to restart costs, and update the framework of Daly (2006) allowing users to specify optimal checkpointing usage intervals that minimize such losses. We derive the machine computational efficiency, which specifies the expected fractional resource allocation that is available for scientific computation. We illustrate the methodology using one year of production runtime data from the \emph{Frontier} supercomputer at Oak Ridge National Laboratory.
\end{abstract}

\section{Introduction}








The study of the effects of faults in shared computing systems, and of resilient computation in the presence of hardware and software faults have been an active area of research for many decades now \cite{young1974first,svobodova1984resilient,zheng2011co,di2019characterizing,taherin2021examining,zhao2022survey}. This research is growing in relevance to scientific high-performance computing (HPC), because with the introduction of the newest Exascale computing platforms has come a reckoning with the fact that applications running at the largest scales face mean times between failure (MTBF) measured in hours, rather than in days or weeks.  Adapting applications and runtime environments to this new landscape of frequent faults is necessarily a high priority for both developers of scientific codes and for HPC facilities.

As of this writing, many datacenters belonging to technology companies are dedicated to training AI models at very large scales, and there is some current literature describing the experiences of such centers in confronting the same issues of frequent computing faults \cite{epoch2024hardwarefailureswontlimitaiscaling,kokolis2025revisiting}.  It is noteworthy, however, that not all these experiences translate exactly into the the types of challenges faced by facilities concerned with scientific HPC.  The reason is that scientific HPC facilities grant time to a far more heterogeneous collection of application codes, including AI loads, but also computational fluid dynamics codes, Earth system model codes, molecular dynamics, cosmology, quantum chemistry, protein folding, and many others (e.g., see \cite{vermascience, e3smfrontier,lammpsdiamond,heitmann2024newworldssimulationslargescale}).  Many of these applications stress computing nodes in different ways, exercising CPU and GPU hardware, memory, interconnects, and other hardware according to a variety of computing patterns, and also requiring differing job sizes (node counts) to run optimally.

Large-scale jobs at a scientific HPC facility are not atypically scheduled for planned runtimes of 24 hours, and face considerable challenges in environments where the MTBF corresponding to their job size is shorter than that time. Designed-in resilience in the face of such fault rates is a simple necessity for such codes, and for the facility itself.  

The most basic and longest-standing measure adopted to ensure resilience is \emph{prophylactic checkpointing}, wherein the code periodically stops the computation in order to write out a summary of its state in sufficient detail that the computation may be resumed from that point in time in the event of a later failure. Checkpoint cadence optimization has been studied in some detail, \cite{young1974first,daly2006higher,epoch2024hardwarefailureswontlimitaiscaling}.  To date, however, this has always been done in a context where the fault rate is considered a constant common to all applications, which is not really the case when application jobs can differ in node count. If we idealize job failures as being largely caused by single-node faults, then failure risk is clearly a function of job size.  In this case, a 2000-node job that runs for one hour is at the same failure risk as a 1000-node job that runs for two hours. The established frameworks for considering optimal checkpointing need to be updated to reflect this reality.

There is a separate issue that is closely tied to the challenges of application failures and resilience: what fraction of a supercomputer's capacity is actually available for scientific computing, after computing operations have been optimally hardened against failures? This is an important question facing HPC facilities, for at least two reasons: it is important to attach realistic performance numbers to HPC platforms, and those numbers are affected by the fraction of FLOPS employed to implement resilience measures rather than on domain science computation; and facilities need to know what fraction of the node-hours allocated to projects are employable for scientific computation, so that they may provision computing resources to projects in a rational manner.

Our contributions to these subjects are twofold. In the first place, we broaden previous work on checkpointing and resilience by changing their operating context from time to \emph{usage} (that is, node-hours). Re-expressing operating costs and failure losses in node-hours rather than merely in computing time allows us to assimilate jobs running at different scales for different times, yielding consistent notions of optimal checkpointing cadence and of the amount of computational resources required to complete a calculation To this end, we introduce the notion of \emph{per-unit-usage failure rate} $R_0$, a parameter with units of failures per node-hour. This parameter is the reciprocal of Mean Usage Between Failures (MUBF), the analog of MTBF in usage space. MUBF is an analogous quantity to "FLOPS per MTBF" \cite{taherin2021examining}. We prefer $R_0$ to MUBF in the technical development that follows below, since it seems to lead to simpler and more natural discussion, but readers more comfortable with MTBF can easily see the connection with ${R_0}^{-1}$.

In the second place, we directly apply the optimal resilience results for applications at HPC facilities to the estimation of \emph{computational efficiency}, which we define as the average fraction of total usage available for practical computation. This is similar in conception to ``Effective Training Time Ratio'' \cite{kokolis2025revisiting}, but differs in the important respect that it is an average over a computational load that is representative of typical scientific HPC loads.

The plan of the rest of the paper is as follows: in $\S\ref{sec:Summary-of-Failure-Related}$ we summarize the mathematical treatment of computing costs incurred as a consequence of a fixed rate of node-local application faults. In $\S\ref{sec:Optimum-Checkpoint-Cadence}$, we discuss the adaptation of classical results on optimal checkpointing from the time domain to the usage domain, from the perspective of domain scientists attempting to complete a specific computation. In $\S\ref{sec:Per-Job-Losses}$, we discuss the costs and losses associated with an application whose computations are partitioned into discrete jobs. In $\S\ref{sec:Machine-Level-Loss}$ we adopt the perspective of an HPC facility running an ensemble of application jobs, and show how the previous results can be used to estimate the computational efficiency of an HPC platform. We discuss our implementation in code of the analysis in $\S\ref{sec:implementation}$. In $\S\ref{sec:Results-From-Frontier}$ we show an example analysis using data from the Oak Ridge National Laboratory machine \emph{Frontier} to furnish a fault rate and an ensemble of jobs characteristic of a scientific HPC facility's computational load. We discuss the implications of our analysis methods and results in $\S\ref{sec:Discussion}$.

\section{Summary of Failure-Related Costs\label{sec:Summary-of-Failure-Related}}

Suppose that a large shared machine runs heterogeneously-sized jobs
concurrently. The machine is composed of a large number of compute nodes which
fail randomly, due to hardware or other system faults. Node faults cause
application jobs to fail at some rate $R_{0}$, measured in failures per unit
usage (i.e. failures/node-hr).  The reciprocal of $R_{0}$ is the mean usage
between failures (MUBF), the usage analog of the mean time
between failures (MTBF) that features in typical discussions of application
failures (e.g. \cite{daly2006higher,young1974first}). MUBF is an analogous quantity to "FLOPS per MTBF" \cite{taherin2021examining}. The reason for working
with rates-per-unit-usage rather than per-unit-time is that failure risk scales
with job size: a 1000-node job that runs for 1 hour is at the same failure risk
from node faults as a 500-node job that runs for two hours.

To protect computations, applications checkpoint at a certain rate.  Since the
failure risk is measured in usage, the checkpoint intervals are also measured
in usage: a 1000-node job runs up failure risk at twice the time rate of a 500-node job, and must accordingly checkpoint twice as frequently. The checkpoint usage
cadence is denoted $u_{c}$ node-hours. This is the amount of usage dedicated to scientific computing between checkpoint writes.

The costs associated with failures and checkpointing are as follows:
\begin{itemize}
\item Each failure costs the loss of all usage consumed by computation or checkpoint writing since the last successful checkpoint;
\item Each checkpoint takes a certain amount of time, and hence consumes
a certain amount of usage;
\item Each restart of a failed job costs a certain amount of usage for setup---including the potential cost of a node failure during startup. 
\end{itemize}
There is an additional cost incurred when a node experiences a non-recoverable hardware
failure and needs to be reset manually or replaced during a maintenance
period. We neglect this cost, since even if it applied to a few nodes
per week, the associated cost in node-hours would be small.

\subsection{Checkpointing Cost}

The principal abatement strategy to protect computations from failures is
\emph{checkpointing}: application codes periodically record their states,
writing it to some non-volatile storage, so that if a failure occurs, only the
output since the last checkpoint is lost. There are two types of checkpoints:
\emph{global checkpoints}, which effectively pause computation until the state
of the application on all nodes is written out to a unitary checkpoint file on a
global storage resource; and \emph{node-local checkpoints}, which save a copy of
the data locally, either to node-local storage or to on-node volatile memory (in
which case a CPU on the node is dedicated to routing the data asynchronously to
global storage), before proceeding with the computation. Clearly, for an
application code to checkpoint locally to volatile memory, it must (a) checkpoint 
amounts of memory small enough to fit, and (b) be explicitly coded to forward its data to the
on-node service responsible for handling checkpoints.

Suppose a job runs on $N$ nodes. Suppose also that each node must checkpoint a
certain amount $M$ of application memory (something like 20\% of available node
memory is typical). Assuming fixed write bandwidth $B_w$ to its storage, this results in a
checkpointing time $\tau_{c}=M/B_w$. Note that this assumes that the entire write bandwidth is available to the node, that is, there is no network contention with other nodes writing their own checkpoints.  Note also the significant approximation that all applications checkpoint the same amount of memory, $M$.

In the case of node-local
checkpointing, the data need only be written out in a time $\tau_{chk}=\tau_c$ to the
local storage, and the computation can resume immediately. In this case, the
usage consumed by the checkpoint is
\begin{equation}
u_{chk}=N\tau_{chk}\textrm{ (node-local storage)}.\label{eq:uchk_nodelocal}
\end{equation}

By contrast, in the case of global checkpointing,
all nodes must ship their data to the storage device, contending for
the available bandwidth to the device, before the computation may
resume. The write time to storage is thus $\tau_{chk}=N\tau_{c}$,
where $\tau_{c}$ is the time required for one node to write out its
checkpoint data. The usage consumed by the checkpoint is
\begin{equation}
u_{chk}=N\times N\tau_{c}=N^{2}\tau_{c}\textrm{ (shared storage)}.\label{eq:uchk_shared}
\end{equation}

The scaling with job size $N$ is significantly different in the two
cases: $\mathcal{O}(N)$ in the case of node-local checkpointing, versus
$\mathcal{O}(N^{2})$ in the case of global checkpointing. Large jobs have
a substantially larger checkpointing usage burden in the shared storage
case than they would in the node-local storage case.

\subsection{Job Failure Cost}

A job that checkpoints each time the application computes for a usage
interval $u_{c}$ runs for a usage $u_{c}+u_{chk}$ between checkpoints.
A job that fails costs a usage $u_{F}$ equal to the elapsed usage
since the last checkpoint, so $0\le u_{F}\le u_{c}+u_{chk}$. The
value of $u_{F}$ is a realization of a truncated exponential random
variable $Y_{F}\sim\textrm{Exp}(R_{0})$ with distribution density
\begin{equation}
\pi_{Y_{F}}(u_{F})=\begin{cases}
\frac{R_{0}\exp\left(-R_{0}u_{F}\right)}{1-\exp\left(-R_{0}(u_{c}+u_{chk})\right)} & 0\le u_{F}\le u_{c}+u_{chk}\\
0 & \textrm{otherwise}.
\end{cases}\label{eq:failure_density}
\end{equation}
The expected cost from such a failure is $\overline{u}_{F}=E_{Y_{F}}(Y_{F})$,
given by
\begin{eqnarray}
\overline{u}_{F} &=&\int_{0}^{u_{c}+u_{chk}}du_{F}\,u_{F}\times\pi_{Y_{F}}(u_{F})\nonumber\\
&=& {R_{0}}^{-1}\left[1-\frac{R_{0}(u_{c}+u_{chk})}{\exp\left(R_{0}(u_{c}+u_{chk})\right)-1}\right]\label{eq:ubar_F}
\end{eqnarray}

\subsection{Restart Cost}

Restarting a job involves reading a checkpoint file in to all the
nodes in a job, and some fixed overhead. Assuming read bandwidth $B_r$ from
storage is related to write bandwidth $B_w$ to storage by $B_w=\alpha B_r$, the cost of a successful
restart is
\begin{equation}
u_{r}=\alpha u_{chk}+\tau_{0}N,\label{eq:success_restart_cost}
\end{equation}
where $\tau_{0}$ is a fixed setup time.

A restart may not be successful, however, since a node failure may
occur while the restart is in process. The probability of a successful
restart is $p_{s}=\exp\left(-R_{0}u_{r}\right)$, and the probability
of a failed restart is $1-p_{s}$. Given that a restart has failed,
the probability density that the failure resulted in a usage loss
$u_{F,r}$ is analogous to Equation (\ref{eq:failure_density}):
\[
\pi_{R}(u_{F,r})=
\begin{cases}
\frac{R_{0}\exp\left(-R_{0}u_{F,r}\right)}{1-\exp\left(-R_{0}u_{r}\right)} &
0\le u_{F,r}\le u_{r}\\
0 & \textrm{otherwise}
\end{cases}.
\]
Analogously to Equation (\ref{eq:ubar_F}), the expected loss due
to a single restart failure is
\[
\overline{u}_{F,r}={R_{0}}^{-1}\left[1-\frac{R_{0}u_{r}}{\exp\left(R_{0}u_{r}\right)-1}\right].
\]
The random variable $N_{F}$ representing the number of restart failures prior to a
successful restart is governed by the geometric distribution, $\textrm{Prob}(N_{F})=(1-p_{s})^{N_{F}}p_{s}$.
By a standard result \cite[Chapter 6]{balakrishnan2003aprimer}, the expected value of $N_{F}$ is
\[
E(N_{F})=\frac{1-p_{s}}{p_{s}}=\exp\left(R_{0}u_{r}\right)-1.
\]

We thus obtain for the expected restart cost
\begin{eqnarray}
\overline{u}_{R} & = & u_{r}+E(N_{F})\times\overline{u}_{F,r}\nonumber \\
 & = & {R_{0}}^{-1}\left[\exp\left(R_{0}u_{r}\right)-1\right].\label{eq:u_R}
\end{eqnarray}
In the limit $R_{0}u_{r}\ll1$, this becomes $\overline{u}_{R}=u_{r}$. 

\section{Optimum Checkpoint Cadence\label{sec:Optimum-Checkpoint-Cadence}}

\begin{figure*}[t]
    \begin{centering}
    \includegraphics[width=\textwidth]{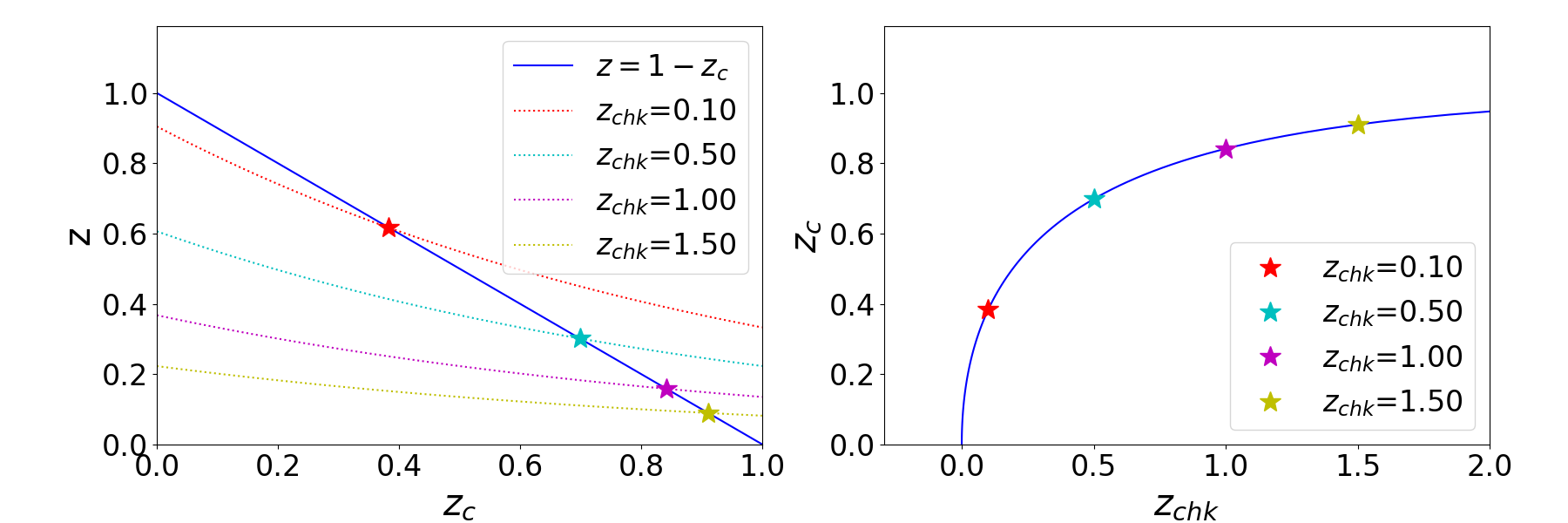}
    \par\end{centering}
    \caption{Optimal checkpointing. The figures show the relation between the dimensionless
    quantities $z_{c}=R_{0}u_{c}$ and $z_{chk}=R_{0}u_{chk}$. The left
    panel illustrates graphically the solution of Equation (\ref{eq:opt_chkpoint})
    for different values of $z_{chk}$, indicated by the star markers. The dotted lines show the right-hand side of Equation (\ref{eq:opt_chkpoint}), whereas the solid line shows the left-hand side of the equation.
    The right panel shows the dependence of $z_{c}$ on $z_{chk},$ with
    the star markers indicating the corresponding solutions from the left
    panel. \label{fig:Optimal-checkpointing}}
\end{figure*}

Consider a project that is expected to require a total usage $U$
to complete its scientific objectives. This will in general require
many restarts to complete. The number of successful checkpoints required
to complete the computation is
\[
N_{S}=\frac{U}{u_{c}}.
\]
In order to complete $N_{S}$ successful checkpoints, the project
will experience $N_{F}$ failures. This is a random variable governed
by the negative binomial distribution, $N_{F}\sim\textrm{NB}(p_{S},N_{S})$,
with probability mass function
\begin{equation}
P(N_{F})=\frac{\left(N_{F}+N_{S}-1\right)!}{N_{F}!\left(N_{S}-1\right)!}{p_{S}}^{N_{S}}\left(1-p_{S}\right)^{N_{F}},\label{eq:negbin}
\end{equation}
where 
\begin{equation}
p_{S}=\exp\left(-R_{0}(u_{c}+u_{chk})\right)\label{eq:p_s}
\end{equation}
is the probability that the computation successfully completes a usage
$u_{c}+u_{chk}$. Equation (\ref{eq:negbin}) can be understood simply,
as the binomial probability of obtaining $N_{F}$ failures and $N_{S}-1$
successes in the first $N_{F}+N_{S}-1$ attempts, multiplied by the
probability of a success on the $(N_{S}+N_{F})$-th attempt.

By a standard result \cite[Chapter 7]{balakrishnan2003aprimer}, the expected number of failures is
\begin{eqnarray*}
\overline{N}_{F}&=&E(N_{F})=\frac{N_{S}(1-p_{S})}{p_{S}}\nonumber\\
&=&N_{S}\left[\exp\left(R_{0}(u_{c}+u_{chk})\right)-1\right].
\end{eqnarray*}

Each failure results in an expected cost that is the sum of the expected
job failure cost and the restart cost. Each success comes at the cost
of a checkpoint write. The total expected cost is thus
\begin{eqnarray*}
L & = & \overline{N}_{F}\left(\overline{u}_{F}+\overline{u}_{R}\right)+N_{S}u_{chk}\\
 & = & U\left[{u_{c}}^{-1}\left(\exp\left(R_{0}(u_{c}+u_{chk})\right)-1\right)\left({R_{0}}^{-1}+\overline{u}_{R}\right)-1\right].
\end{eqnarray*}

The optimal $u_{c}$ is obtained by minimizing this cost with respect
to $u_{c}$:

\begin{eqnarray*}
0 & = & \frac{dL}{du_{c}}\\
 & = & U\left({R_{0}}^{-1}+\overline{u}_{R}\right){u_{c}}^{-2}\nonumber\\
 &&\times\left[1-\left(1-R_{0}u_{c}\right)\exp\left(R_{0}(u_{c}+u_{chk})\right)\right],
\end{eqnarray*}
from which we obtain the relation
\begin{equation}
1-R_{0}u_{c}=\exp\left(-R_{0}(u_{c}+u_{chk})\right).\label{eq:opt_chkpoint}
\end{equation}
This formula is analogous to Equation (21) of \cite{daly2006higher},
with the difference that costs here are measured in usage (node-hours)
rather than in time. We may solve Equation (\ref{eq:opt_chkpoint})
for $u_{c}$ numerically by a 1-dimensional root-find. Setting $z_{c}\equiv R_{0}u_{c}$,
$z_{chk}\equiv R_{0}u_{chk}$, the solution is illustrated in the
two panels of Figure \ref{fig:Optimal-checkpointing}.

Once $z_{c}$ is determined as illustrated in the Figure, we obviously
have $u_{c}=z_{c}/R_{0}$. There are two instructive limits: When
$R_{0}\rightarrow0$, we have $z_{c}\rightarrow0$ and $z_{chk}\rightarrow0$,
and Equation (\ref{eq:opt_chkpoint}) is approximately
\[
1-z_{c}\approx1-z_{c}-z_{chk}+\frac{1}{2}\left(z_{c}+z_{chk}\right)^{2},
\]
so that 
\[
z_{c}\approx\sqrt{2z_{chk}}-z_{chk}\approx\sqrt{2z_{chk}},
\]
to leading order in $z_{chk}$. Hence we have
\begin{equation}
u_{c}\approx\sqrt{\frac{2u_{chk}}{R_{0}}}
\label{eq:young}
\end{equation}
which is the well-known first-order result of \cite{young1974first},
translated from time to usage. The checkpoint cadence thus gets larger
as the failure rate decreases, as accords with intuition.  Another noteworthy
property of this limiting case is that if checkpointing is performed to a
central storage server, then according to Equation (\ref{eq:uchk_shared}) we have
$u_{chk}\propto N^2$, and $u_c\propto N$. Hence the checkpointing \emph{time cadence}
$t_{chk}=u_{chk}/N$ tends to a constant independent of $N$ in this limit.

In the opposite limit $R_{0}\rightarrow\infty$, it is clear that
$z_{chk}\rightarrow\infty$, $z_{c}\rightarrow1$. We may rewrite
Equation (\ref{eq:opt_chkpoint}) as
\[
(1-z_{c})=\exp\left(1-z_{c}\right)\times\exp\left(-1-z_{chk}\right),
\]
 and expanding the first exponential we get
\[
z_{c}\approx1-\exp\left(-1-z_{chk}\right),
\]
which is to say
\[
u_{c}\approx{R_{0}}^{-1}-{R_{0}}^{-1}\exp\left(-1-R_{0}u_{chk}\right).
\]
 Thus, as the failure rate increases, the optimal checkpoint time
approaches ${R_{0}}^{-1}$, the mean usage between failures.

\section{Per-Job Losses\label{sec:Per-Job-Losses}}

We determined the optimal checkpoint cadence $u_{c}$ by considering
usage from the point of view of a single project, according to a model
in which the project simply restarts every failed job until it reaches
its required computational usage. On real-world supercomputers, applications
compute in discrete units of jobs, each confined to some number $N$
of nodes and scheduled for some time $T$ (typically limited to 12
or 24 hours). We will refer to such a job as being of ``type'' $(N,T)$.
Within that scheduled usage, the application checkpoints at its optimal
rate for prophylactic purposes, i.e., to abate loss. If the job succeeds,
it writes the final checkpoint before exiting, so that it may restart
later from that stage of the computation. Each of the prophylactic
checkpoints takes up a usage $u_{chk}$ which counts as a loss, since
on a fault-free system they would be unnecessary. The final checkpoint
would be required even on a fault-free system, and therefore does
not count as a loss.

Suppose that a project requires $N_{J}$ jobs of type $(N,T)$ to reach its
computational objective. All jobs incur checkpointing costs irrespective of
whether they succeed or fail. If the failure probability of a job is
$P(\textrm{Failure})$, then we expect $N_{J}\times P(\textrm{Failure})$ failed
jobs, each of which will incur usage costs: besides the checkpoint cost, failed
jobs incur failure costs, amounting in expectation to $N_{J}\times
P(\textrm{Failure})\times\overline{u}_{F}$, as well as restart costs, amounting
in expectation to $N_{J}\times P(\textrm{Failure})\times \overline{u}_{R}$. The sum of
these costs is the computational loss due to machine instability to jobs of type
$(N,T)$. Dividing this cost by $N_{J}$ (an uninteresting normalization) gives
the cost per job.

\subsection{Expected Per-Job Checkpointing Loss}

A successful job of type $(N,T)$ will comprise $N_{chk}$ checkpoints,
including the final, non-prophylactic checkpoint, where
\begin{equation}
N_{chk}=\left\lceil \frac{NT}{u_{c}+u_{chk}}\right\rceil.
\label{eq:Nchk}
\end{equation}
The notation $\lceil x\rceil$ denotes the smallest integer larger
than $x$, i.e. the ``round up'' function.

The probability that an individual checkpoint is successfully written
is $p_{S}$, given in Equation (\ref{eq:p_s}). Let us define the
random variable $K$, representing the number of successful checkpoints
prior to failure. $K$ takes values $k=0,\ldots,N_{chk},$where $K=N_{chk}$
is the event ``Success'' and $K<N_{chk}$ is the event ``Failure''.
We have
\begin{eqnarray*}
P\left(\textrm{Success}\right) & = & p_{S}^{N_{chk}}\\
P\left(\textrm{Failure}\right) & = & 1-p_{S}^{N_{chk}}.
\end{eqnarray*}
$K$ has a truncated geometric distribution, with probability mass
function
\[
P\left(K=k\right)=\begin{cases}
p_{S}^{k}(1-p_{S}) & k<N_{chk}\\
p_{S}^{N_{chk}} & k=N_{chk}.
\end{cases}
\]
It is straightforward to verify that $\sum_{k=0}^{N_{chk}}P(K=k)=1$
using the formula for summation of a geometric series.

The loss to checkpointing for a single job of type $(N,T)$ with $K=k$
is
\[
L_{chk}(N,T,k)=\begin{cases}
ku_{chk} & k<N_{chk}\\
(N_{chk}-1)u_{chk} & k=N_{chk},
\end{cases}
\]
since we don't count a checkpoint written at $K=N_{chk}$ as a loss.
The per-checkpoint usage $u_{chk}$ is given by Equation (\ref{eq:uchk_nodelocal})
or (\ref{eq:uchk_shared}), depending on the checkpointing strategy
(node-local or global). 

The expected per-job loss to checkpointing is therefore
\begin{align}
\overline{L}_{chk}(N,T) &=  u_{chk}\Bigg[\sum_{k=0}^{N_{chk}-1}kP\left(K=k\right)\nonumber\\
&\hspace{1cm}+(N_{chk}-1)P\left(K=N_{chk}\right)\Bigg]\nonumber\\
&=  u_{chk}\Bigg[(1-p_{S})\sum_{k=0}^{N_{chk}-1}kp_{S}^{k}\nonumber\\
&\hspace{1cm}+(N_{chk}-1)p_{S}^{N_{chk}}\Bigg].
 \label{eq:LBar_1}
\end{align}
The sum in this expression may be evaluated by recognizing that it is proportional to the derivative with respect to $p_S$ of the sum of a geometric series, $\sum_kk(p_S)^k=p_S\frac{d}{dp_S}\sum_k(p_S)^k$. We find
\begin{equation}
\overline{L}_{chk}(N,T) = u_{chk}\frac{p_{S}-p_{S}^{N_{chk}}}{1-p_{S}}.\label{eq:LBar_2}
\end{equation}

\subsection{Total Expected Per-Job Loss}

The per-job loss to failure is
\begin{equation}
\overline{L}_{F}(N,T)=P\left(\textrm{Failure}\right)\times\overline{u}_{F}=\left(1-p_{S}^{N_{chk}}\right)\overline{u}_{F},\label{eq:Lbar_chk}
\end{equation}
where $\overline{u}_{F}$ is given by Equation (\ref{eq:ubar_F}).

The per-job loss to restart is
\begin{equation}
\overline{L}_{R}(N,T)=P\left(\textrm{Failure}\right)\times \overline{u}_{R}=\left(1-p_{S}^{N_{chk}}\right)\overline{u}_{R},\label{eq:Lbar_R}
\end{equation}
where $\overline{u}_{R}$ is given by Equation (\ref{eq:u_R}).

The total expected per-job loss is
\[
\overline{L}(N,T)=\overline{L}_{chk}(N,T)+\overline{L}_{F}(N,T)+\overline{L}_{R}(N,T).
\]

\subsection{Expected Per-Job Usage}

In the computation of machine efficiency, we will also require the
expected usage of a job. This is not the same as its scheduled usage,
because of the possibility of job failure.

The usage of a job of type $(N,T)$ with $K=k$ is
\[
U(N,T,k)=\begin{cases}
k\times(u_{c}+u_{chk})+u_{F,k+1} & k<N_{chk}\\
N_{chk}(u_{c}+u_{chk}) & k=N_{chk}
\end{cases},
\]
where $u_{F,k+1}$ is the random failure loss in the $(k+1)$-th usage
interval. We already know that $E\left(u_{F,k+1}\right)=\overline{u}_{F}$,
given by Equation (\ref{eq:ubar_F}). We take the expectation of $U(N,T,k)$:
\begin{eqnarray}
\overline{U}(N,T) & = & \sum_{k=0}^{N_{chk}-1}P(K=k)\times k\times(u_{c}+u_{chk})
\nonumber\\
&&+P\left(\textrm{Failure}\right)\overline{u}_{F}\nonumber\\
&&+P\left(\textrm{Success}\right)N_{chk}(u_{c}+u_{chk})\nonumber \\
 & = & \frac{p_{S}(1-p_{S}^{N_{chk}})}{1-p_{S}}(u_{c}+u_{chk})+(1-p_{S}^{N_{chk}})\overline{u}_{F}.\label{eq:expected_usage}
\end{eqnarray}

\section{Machine-Level Loss\label{sec:Machine-Level-Loss}} A supercomputer will
run a mix of different jobs of different types during a given period. This could
be described by a schedule $(N_{l},T_{l})$, $l=1,\ldots,N_{Sched}$ of
$N_{Sched}$ jobs, with the $l$-th such jobs described by the type
$(N_{l},T_{l})$. This formulation might be appropriate if one is estimating loss
for a production machine based on past usage data, or for a new machine based on
scaled-up jobs from an existing production machine. Alternatively, a
supercomputer facility may use its annual allocation programs to get an idea of
a distribution $\rho(N,T$) representing the number of jobs expected to run in a
bin of width $(\Delta N,\Delta T)$ centered on ($N,T)$ on a production machine
during (say) a fiscal year. Either of these formulations may be used to estimate
expected loss. Here we adopt the schedule approach. Replacing a schedule with a
binned distribution is straightforward, and we will not belabor the required
changes in formulas here.

A schedule $(N_{l},T_{l})$, $l=1,\ldots,N_{Sched}$ represents a
total expected usage
\[
\overline{U}_{Tot}=\sum_{l=1}^{N_{Sched}}\overline{U}(N_{l},T_{l}).
\]

The expected loss during the course of this schedule is
\begin{align}
\overline{L}&=\sum_{l=1}^{N_{Sched}}\overline{L}(N_{l},T_{l})\nonumber\\
&=\sum_{l=1}^{N_{Sched}}\left[\overline{L}_{chk}(N_{l},T_{l})+\overline{L}_{F}(N_{l},T_{l})+\overline{L}_{R}(N_{l},T_{l})\right].\nonumber
\end{align}
The remaining usage, $\overline{U}_{Tot}-\overline{L}$ is actual
scientific computation. The \emph{efficiency} $e$ of the machine
over the course of this schedule is
\[
e\equiv\frac{\overline{U}_{Tot}-\overline{L}}{\overline{U}_{Tot}}=1-\frac{\overline{L}}{\overline{U}_{Tot}}.
\]
 This is the key parameter to be recovered by this exercise. It tells
us the degree to which allocations must be ``stretched'' in order
to accomplish their required usage: If a project requires usage $U$
node-hours to accomplish their science, the required allocation $U_{Alloc}$
satisfies $U=eU_{Alloc}$, so that
\begin{equation}
U_{Alloc}=\frac{U}{e}.
\label{eq:overprovision}
\end{equation}
A perfect machine has $e=1$, and allocates $U_{Alloc}=U$. As the
efficiency decreases, the required allocation increases. If over a
period of time (e.g. a fiscal year) the maximum available usage is
$U_{max}$, then the usage available for scientific computation is
$eU_{max}$. 

The specific losses that contribute to this efficiency are also of interest. We define
\begin{eqnarray}
l_{chk}&\equiv&\frac{\sum_{l=1}^{N_{Sched}}\overline{L}_{chk}(N_{l},T_{l})}{\overline{U}_{Tot}},\label{eq:lchk_rel}\\
l_F&\equiv&\frac{\sum_{l=1}^{N_{Sched}}\overline{L}_{F}(N_{l},T_{l})}{\overline{U}_{Tot}},\label{eq:lF_rel}\\
l_R&\equiv&\frac{\sum_{l=1}^{N_{Sched}}\overline{L}_{R}(N_{l},T_{l})}{\overline{U}_{Tot}}.\label{lR_rel}
\end{eqnarray}

Then we obviously have $e=1-l_{chk}-l_F-l_R$.

\section{Implementation\label{sec:implementation}}

We have implemented the above analysis as a small collection of Python functions, which are invoked by a Jupyter notebook to produces the results that we present below. The code is in a publicly available \emph{GitHub} repository at \url{https://github.com/CarloGraziani/HPC_Computational_Efficiency}. The repository also makes available the scheduling data from the \emph{Frontier} supercomputer, described in $\S$\ref{sec:Results-From-Frontier}, which we used to create a schedule $(N_{l},T_{l})$, $l=1,\ldots,N_{Sched}$ as described in $\S$\ref{sec:Machine-Level-Loss}.

The code calculates an optimal checkpointing usage cadence for each individual job, and calculates the resulting overall losses and machine efficiency that result from the assumption that all jobs are checkpointed at that rate.

One subtlety that we noticed in the process of analyzing the data is that the job schedule distribution is strongly-clustered in job sizes and job durations, and this fact interacts with the discrete truncation of the per-job checkpoint number $N_{chk}$ given in Equation (\ref{eq:Nchk}) to produce slight non-smooth artifacts in the efficiency and loss curves shown in Figures \ref{figure:checkpoint_cadence}, \ref{fig:e_vs_R0}, and \ref{fig:e_vs_bandwidth}. The code offers an option to compute $N_{chk}$ as a continuous function (by refraining from rounding up), which is possible because the loss formulas in $\S$\ref{sec:Per-Job-Losses} work irrespective of whether or not $N_{chk}$ is integer-valued. Triggering this option removes the artifacts.  We do not avail ourselves of that choice below, preferring instead to explain their origin while keeping the (more correct) formulation in terms of an integer-valued $N_{chk}$.

\section{Results From \emph{Frontier} Usage Data\label{sec:Results-From-Frontier}}

We now describe the application of this framework for analyzing machine efficiency to data from the \emph{Frontier} supercomputer. After a description of the machine and a discussion of the data and model parameters, we provide analyses of computational efficiency, of the effect of suboptimal checkpointing, and of the influence on efficiency of job size and duration, of application failure rate, and of bandwidth to storage.

\subsection{Computational Efficiency of \emph{Frontier}}

Frontier is an exascale supercomputer maintained at the
Oak Ridge Leadership Computing Facility (OLCF). Frontier is currently Number 2 on November 2024’s Top500 list with a High-Performance Linpack (HPL) score
of 1.353 EFlop/s. The system is comprised of 9,408 HPE
Cray EX235a nodes, each with one 64-core AMD EPYC
7A53  CPU and four AMD
MI250X GPUs. Each MI250X contains 2 Graphics Compute Dies (GCDs), each of which appears as a single logical device for most applications. Each compute node has
512 GB of DDR4 memory and 512 GB of high-bandwidth
memory (HBM2E), 64 GB per GCD. The CPU is connected
to the GPUs via AMD’s Infinity Fabric which delivers a
bandwidth of 36+36 GB/s. All GCDs on a Frontier node are
interconnected via Infinity Fabric delivering up to 50+50 GB/s
for GCDs across GPUs, and up to 200+200 GB/s for GCDs on
the same GPU. Compute nodes on Frontier are interconnected
via HPE’s Slingshot 11 interconnect.

\subsection{System Parameters\label{subsec:system-parameters}}

\subsubsection{Application Failure Rate\label{subsubsec:failure_rate}}
Node failures on \emph{Frontier} are detected and logged, and are found to occur almost exclusively in conjunction with the failure of applications that are utilizing them at the time of failure---that is, nodes unstressed by an application rarely fail spontaneously. We therefore estimate the parameter $R_0$ based on the node failure rate. In this way we obtain the estimate $R_0=2.0\times 10^{-5}~\textrm{(node-hours)}^{-1}$. The resulting failure rate is necessarily a slight underestimate of the true $R_0$, because a job may occasionally also fail due to some system issue that does not cause a node failure. We neglect this error in this study.

\subsubsection{Storage Bandwidth}
\emph{Frontier} utilizes a central storage server mounting a \emph{Lustre} filesystem (dubbed Orion) composed of a Flash-based performance tier of 5,400 nonvolatile memory express (NVMe) devices providing 11.5 petabytes (PB) of capacity at peak read-write speeds of 10 TB/s. In practice, most jobs see 
equal read and write bandwidths of roughly $B_w=B_r=6~\textrm{TB~s}^{-1}$. Since read and write bandwidths are equal, we set the parameter $\alpha$ in Equation (\ref{eq:success_restart_cost}) to $\alpha=1$.  We assume an average checkpoint size of 20\% of on-node memory, amounting to $M=200\textrm{~GB}$. This leads to a per-node checkpoint time $\tau_c=9.26\times 10^{-6}\textrm{~hrs}$. In practice, of course, different applications may checkpoint different amounts of memory, so that the assumption of a shared value of $M$ is a significant approximation.

\subsubsection{Fixed Setup Time}
On \emph{Frontier}, the setup time $\tau_0$ in Equation~(\ref{eq:success_restart_cost}) is $2\textrm{s}$. This is a negligible term, which we include in our analysis for the sake of completeness.

\subsection{Job Schedule\label{subsec:schedule}}

\begin{figure}[t]
\begin{centering}
\includegraphics[width=0.5\textwidth]{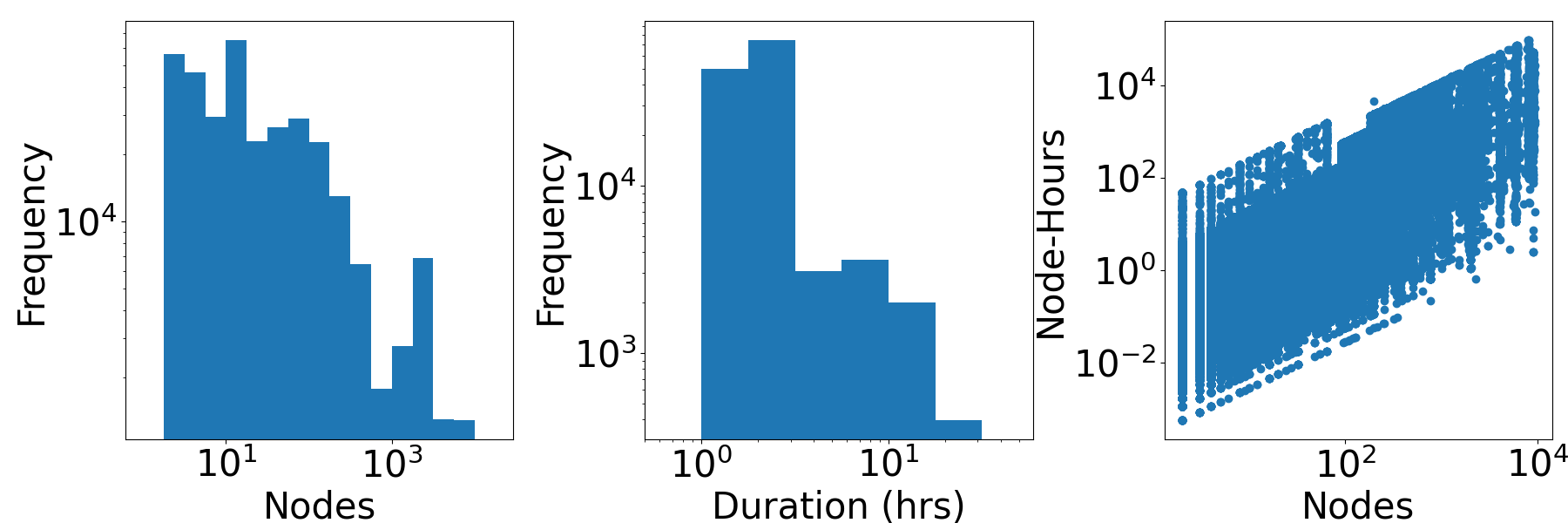}\par
\end{centering}
\caption{Distribution of job characteristics from \emph{Frontier} used in this study.}\label{figure:job_dist}
\end{figure}

We use a database of 331,640 jobs from one year (CY 2024) of operations on \emph{Frontier}. The job selection criteria used for the analysis required that the jobs examined be from one of the facility's two major allocation programs, the Innovative and Novel Computational Impact on Theory and Experiment (INCITE) program and the ASCR Leadership Computing Challenge (ALCC) program.  In addition, jobs that were run on a single node were excluded, since such jobs are very typically interactive or small debugging runs. The job characteristics are shown in Figure~\ref{figure:job_dist}.

Each job is characterized by its size $N_l$ in nodes, its requested time, and its actual runtime.  Many jobs ran for an actual runtime that was substantially less than its requested time: the ratio of mean runtime to mean requested time is $r=0.46$. Since most jobs terminate normally, this ratio is indicative of conservative scheduling by users rather than of a high failure rate.  This presents a problem for our analysis, because the requested times $T_l$ are not really representative of the time required for successful job completion.  We account for this fact by multiplying all requested times by the ratio $r$, so as to adjust the requested times to be closer to what the applications actually required to complete a job.

Note that we do not use (or even know) the termination status of any of these jobs. We only use this data for the purpose of establishing a plausible distribution of job sizes and requested durations that is representative of a realistic computational load due to scientific applications running on \emph{Frontier}.

\subsection{Computational Efficiency of \emph{Frontier}}

Based on the parameters given in $\S$\ref{subsec:system-parameters} and the job schedule described in $\S$\ref{subsec:schedule}, we find that \emph{Frontier} has a computational efficiency $e=0.957$, due to costs and losses $l_{chk}=0.012$, $l_F=0.028$, $l_R=0.003$.  The high efficiency on \emph{Frontier} suggests that if users on the machine in fact checkpoint optimally, the OLCF need only overprovision projects  with node-hours by a fractional amount $e^{-1}-1=4.5\%$ compared to their approved allocation requests.

\subsection{Importance of Proper Checkpointing\label{subsec:importance_checkpointing}}

\begin{figure}[h]
\begin{center}
\includegraphics[width=0.5\textwidth]{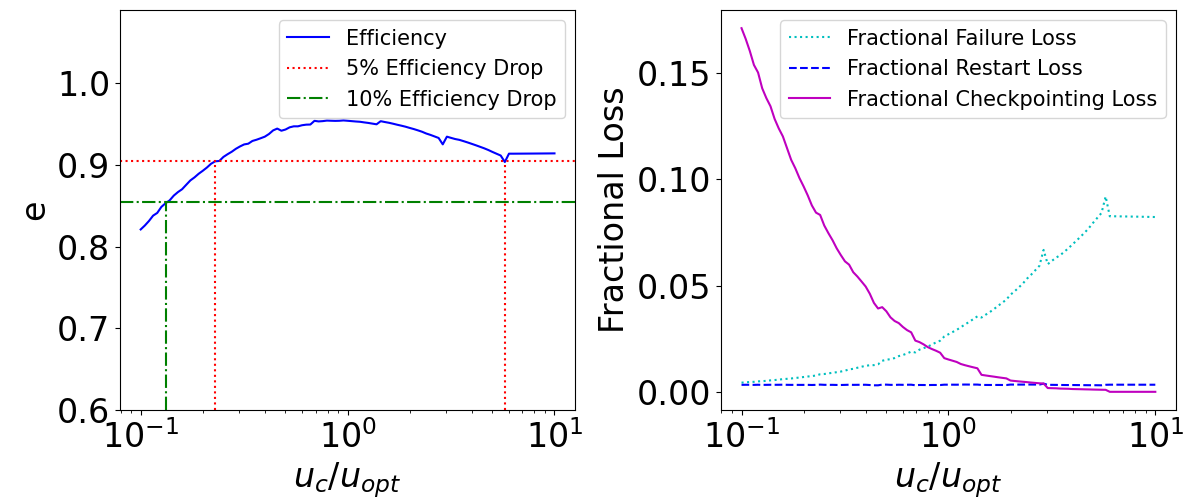}
\caption{Effect of suboptimal checkpoint interval selection on efficiency. The $x$-axis shows the common fraction by which all checkpoint intervals are shifted from their respective optimal value.\label{figure:checkpoint_cadence}}
\end{center}
\end{figure}

The computational efficiency given above corresponds to a situation in which all applications checkpoint at their own optimal rate.  We now investigate the effect of suboptimal checkpointing on efficiency.

In the first place, what happens if applications on \emph{Frontier} do not do any prophylactic checkpointing at all? To answer this question, we repeat the analysis while setting the checkpointing interval $u_{chk}$ of each job equal to its total requested usage. The result is that the computational efficiency of \emph{Frontier} would then be $e=0.914$, a drop of 4.3\% from its optimal efficiency.  For reference, with this efficiency projects would need to be overprovisioned with node-hours by a factor of 9.4\% in order to achieve their requested targets on node-hours used for science computations.

In Figure \ref{figure:checkpoint_cadence} we show the effect of shifting checkpointing cadence of all jobs by a common factor. The left panel of the figure shows efficiency as a function of checkpointing cadence, while the right panel shows the dependence on checkpointing cadence of the individual loss terms. The left panel of the figure shows the points at which the efficiency drops by 5\% and by 10\% from peak. If the checkpoint interval is displaced from optimality by a factor of 5 or so in either direction, the result is a loss of efficiency of 5\%, and if the cadence is about a factor of 10 too frequent,  the loss can be as large as 10\%.  Despite appearances, these kinds of inefficiencies are not implausible mistakes to make. Consider, for example an $N=2000$ node job on \emph{Frontier}, which, according to Equation (\ref{eq:uchk_shared}) has a checkpointing usage $u_{chk}=37.04~\textrm{node-hours}$. Since $z_{chk}=R_0u_{chk}\ll 1$, we may use the formula in Equation (\ref{eq:young}) to compute the checkpoint usage cadence $u_c=1925~\textrm{node-hours}$, and hence a time cadence $u_c/N\approx 1~\textrm{hr}$.  Hence, checkpointing at a time rate of every 12 minutes, or every 5 hours, would lead to an efficiency degradation of 5\% for this type of job.

\subsection{Efficiency Versus Job Size and Duration}

\begin{figure}[t]
\begin{center}
\includegraphics[width=0.5\textwidth]{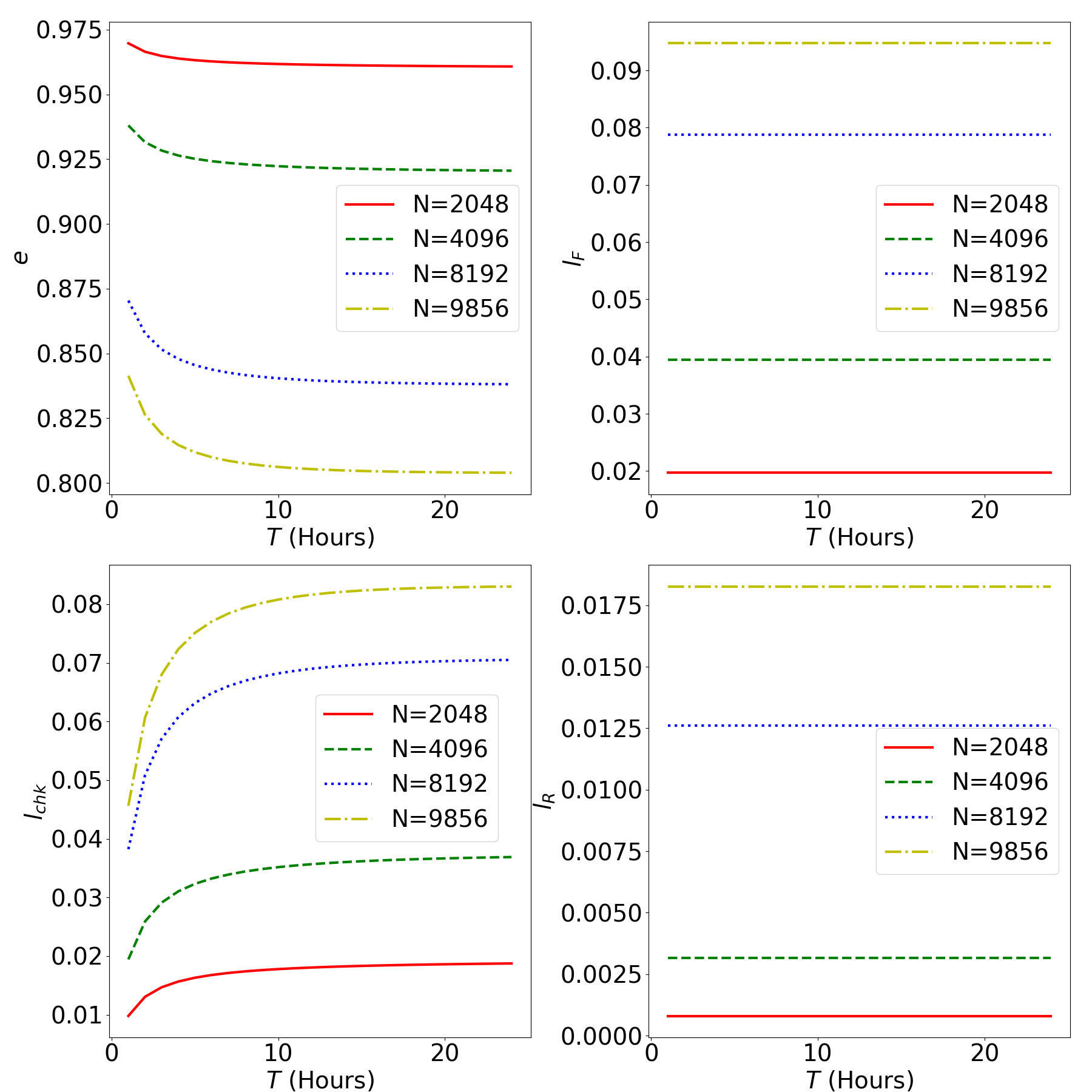}
\caption{Efficiency and relative losses as a function of requested duration.\label{figure:eff_loss}}
\end{center}
\end{figure}

Figure \ref{figure:eff_loss} shows the efficiency and relative losses expected for jobs of different node sizes and requested durations.  Efficiency is shown in the top-left panel, while the loss terms due to faults, checkpointing, and startup are shown in the remaining panels. Quite generally, we see that the efficiency declines with increasing requested duration, saturating at large durations due to the saturation of checkpointing cost.  The latter occurs because as duration becomes large so does $N_{chk}$, and it is clear from Equations (\ref{eq:LBar_2}), (\ref{eq:expected_usage}), and (\ref{eq:lchk_rel}) that in this limit the checkpointing loss saturates. Intuitively, this is explainable in terms of the increasing certainty of a failure with increasing duration, which leads to a finite number of expected checkpoint writes during the course of a finite expected usage.

The losses $l_F$ and $l_R$ due respectively to failures and restarts are constants as a function of requested job duration. Algebraically this is because, according to Equations (\ref{eq:Lbar_chk}), (\ref{eq:Lbar_R}), and (\ref{eq:expected_usage}), the quantities $\overline{L}_F(N,T)$, $\overline{L}_R(N,T)$, and $\overline{U}(N,T)$ all depend on duration $T$ through the common multiplicative factor $(1-{p_S}^{N_{chk}})$, which therefore cancels in the computation of $l_F$ and $l_R$ according to Equations (\ref{eq:lF_rel}--\ref{lR_rel}).

\subsection{Efficiency Versus Failure Rate}

\begin{figure}[h]
\begin{center}
\includegraphics[width=0.5\textwidth]{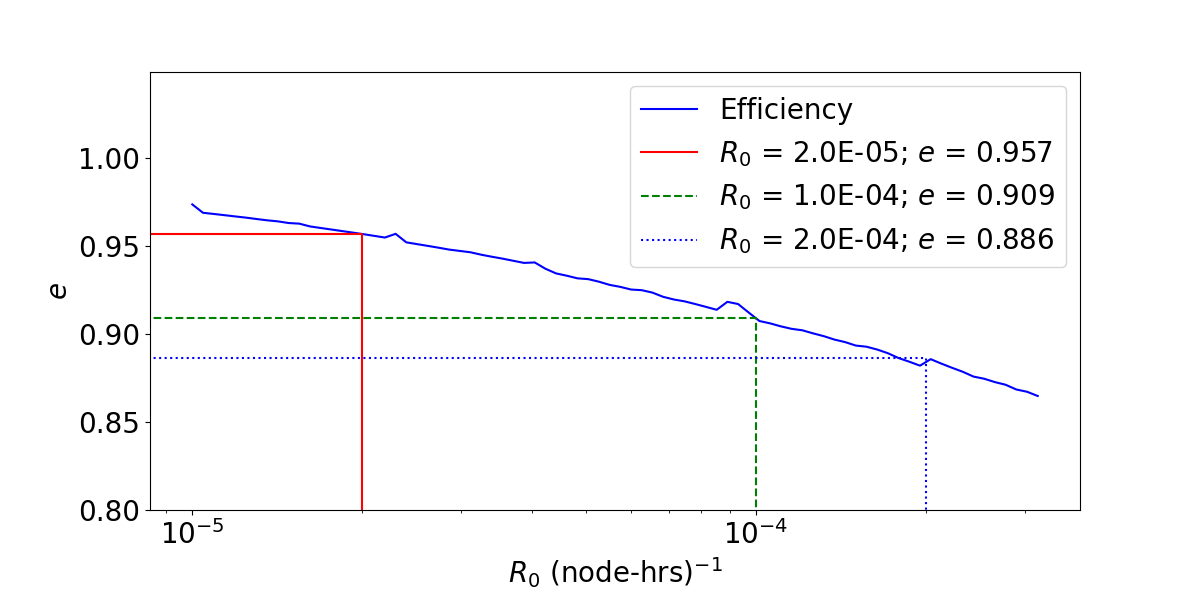}
\caption{Efficiency as a function of failure rate.\label{fig:e_vs_R0}}
\end{center}
\end{figure}

Figure \ref{fig:e_vs_R0} displays the decline of efficiency with failure rate. As indicated in the figure, the efficiency drops from 95.7\% when $R_0=2\times 10^{-5}$ to about 90.9\% when $R_0=10^{-4}$, and to 88.6\% when $R_0=2\times 10^{-4}$.  These latter rates are shown for the purpose of illustrating how a new machine's efficiency might be expected to improve as it progresses in stability from the early stages of deployment to mature operations, during which time failure modes are identified and abated.  Note also that these rates are quite sensitive to the bandwidth available for transferring data to storage, as we discuss next.

\subsection{Efficiency Versus Bandwidth to Storage}

\begin{figure}[b]
\begin{center}
\includegraphics[width=0.4\textwidth]{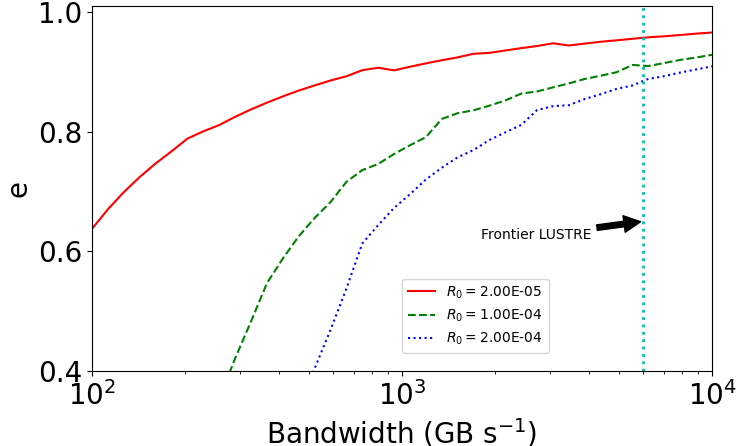}
\caption{Efficiency as a function of bandwidth to storage.\label{fig:e_vs_bandwidth}}
\end{center}
\end{figure}

Figure \ref{fig:e_vs_bandwidth} displays the dependence of efficiency on bandwidth to storage, assuming a shared storage device.  It is clear from the figure that bandwidth is a critically important consideration when designing a machine that is to support resilient computation in the presence of frequent machine faults, because there is a striking drop in efficiency with decreasing bandwidth. This degradation of efficiency is even more striking for the higher failure rates displayed in the figure. This is largely a consequence of the $\mathcal{O}(N^2)$ scaling of $u_{chk}$ that occurs for checkpointing to shared storage in Equation (\ref{eq:uchk_shared}), which severely penalizes checkpointing for large jobs. Note that we can partly understand the effect of application-specific variations in the checkpointable memory $M$ by treating it as equivalent to application-specific bandwidth to storage with fixed, shared $M$. Figure \ref{fig:e_vs_bandwidth} may be viewed as illustrating the fact that efficiency can be very sensitive to amount of checkpointable memory, especially at higher failure rates.

\section{Discussion\label{sec:Discussion}}


We have presented a framework for evaluating the computational efficiency of a scientific HPC machine in the presence of frequent node faults. The chief novelty here is the transposition of some previous results on checkpoint cadences \cite{daly2006higher,young1974first} from the time domain to the usage domain, as this is more relevant to the situation facing modern scientific HPC, because of the heterogeneity in job sizes and durations that is characteristic of such platforms.

We can anticipate that future machines that move beyond the current Exascale class will experience even higher failure rates per node-hour than the current generation, due to the growing complexities of node design and of system software, and to growing transistor counts on nodes.  This observation leads to some considerations that seem relevant to the design of such machines, and of their software stacks.

In the first place, it seems likely that checkpointing will remain the principal application failure abatement strategy in future.  While there exist checkpoint-free resilience strategies \cite{chen2011algorithm},  they principally apply to single solver algorithms, and would be challenging to apply to the kinds of multi-physics, multi-solver codes that typify scientific HPC. This being the case, it seems clear to us that it is imperative that future machines should be designed so as to avoid the $\mathcal{O}(N^2)$ scaling of usage consumed by checkpoint reads and writes described by Equation (\ref{eq:uchk_shared}), because that scaling is likely to become increasingly toxic with increasing node counts and fault rates.  Node-local storage seems like an obvious design strategy to avoid this scaling. Alternatively, on-node memory and hardware specialized for the purpose of receiving checkpointable data and transmitting it to central storage while the computation proceeds on CPUs and GPUs could be a viable choice.

Future machines could also provide explicit system support for checkpointing, making the process more standardized and requiring less specialized application-level infrastructure development by science application developers.  Some thought could be given to resilient strategies for system-level dynamic deletion of older checkpoints to make space for new ones, a consideration of increasing importance as checkpointable application memory seems likely to grow with machine size.

In addition, better facilities for resilient computing could be supplied, so that an application that suffers a node failure does not necessarily exit immediately and go back to the queue, but rather can restart gracefully from the last checkpoint and complete the computations of its planned job duration.  While this sort of resilience is currently easy to obtain for some applications, such as AI loads that merely lose an inessential minibatch's worth of gradient components when a node is lost \cite{epoch2024hardwarefailureswontlimitaiscaling}, other applications require more resilience engineering at present.  For example, a computational fluid dynamics simulation or an Earth system model cannot survive the loss of the part of their computational domain lost to a node failure without rolling back to the previous checkpoint.  Job queuing software could be designed to recognize application failures due to node faults, and to provision replacement nodes and facilitate replacement MPI communicators (for example) which allow the application to recover gracefully from the last checkpoint.  Replacement nodes could be pre-allocated to job partitions, based on the expected number of node faults during each job.

In this work, we have used a single failure rate $R_0$ as characteristic of all applications running on a machine. However, it seems very likely that different types of applications fail at different rates, depending on the manner in which they stress the node hardware and/or trigger problems due to system software.  It would benefit individual application developers to establish estimates of their applications' individual failure rates, because this could lead to more refined estimates of their optimal checkpointing cadence, with measurable benefits for their computational efficiency. Incorporation of application-specific checkpointable memory $M$ into the formalism for computing optimal checkpoint intervals would also bring substantial benefits, because this is analogous to varying the bandwidth to storage for fixed $M$, and as Figure \ref{fig:e_vs_bandwidth} illustrates, efficiency can be a very sensitive function of bandwidth, especially at higher failure rates.

From the perspective of HPC facilities, it seems extremely desirable that node fault rates and application failure rates should be measured and monitored over time.  This is not an easy task, because the relevant log files accumulate vast amounts of data, most of which is irrelevant to this task, and a good deal of it can wind up in different data silos: kernel messages, application standard error files, and scheduler logs are an example of this. Moreover, some of the highest-quality information concerning application faults is in application output files, which application teams do not necessarily share by default with the facility.  And, unfortunately, job exit status can be an unreliable indicator of job success or failure, depending on the design of the job submission script.  It would be well-worth designing instrumentation for capturing and identifying such faults and failures with good specificity and sensitivity, and building it into the machine from the outset, rather than relying on clever datamining of logfiles.

Good capture of failure rates, possibly broken out by application type, would benefit HPC facilities by allowing them to accurately overprovision node-hours to application teams by factors suggested by the relevant computational efficiencies.  This overprovisioning, described by Equation (\ref{eq:overprovision}) would simplify the transactions between facilities and application teams, since the latter would not require compensating node-hours for each failed job, having been, in a sense, ``pre-compensated'' for machine faults by the expected amount.  For this to work well, HPC facilities would certainly require accurate metering of fault and failure rates. 

One important gap in this work is that we have only considered node-local application failures here.  These appear to constitute the majority of failures, but do not make up the whole story.  Some failures are collective, such as slowdowns or hangs due to network congestion.  In addition, currently an application that hangs rather than exiting cleanly consumes usage to no purpose until it is explicitly killed.  This is an effect that could be modeled, but of course a better solution would be platform-level software (for example in the scheduler) to recognize hangs and stop the offending jobs.

\section*{Acknowledgments}
We gratefully acknowledge the use of the resources of the Argonne Leadership Computing Facility at Argonne National Labratory. This work was supported by the Office of Advanced Scientific Computing Research, Office of Science, U.S. Department of Energy, under Contract DE-AC02-06CH11357. This research used resources of the Oak Ridge Leadership Computing Facility at the Oak Ridge National Laboratory, which is supported by the Office of Science of the U.S. Department of Energy under Contract No. DE-AC05-00OR22725.
We are grateful to Nwamaka Okafor (ANL/ALCF), Shilpika (ANL/ALCF), and Scott Simmerman (ORNL/OLCF) for valuable
discussions about log files and failure detection and to Mike Papka,
Venkatram Vishwanath, Eric Pershey, Susan Coghlan, and Michael Zhang (ANL/ALCF) for discussion
of issues raised by application failures for HPC facilities.


\bibliographystyle{siamplain}
\bibliography{refs}

\end{document}